\documentstyle[12pt,subeqnarray,amssymbols,epsf]{article}

\def\beq{\begin{equation}}
\def\eeq{\end{equation}}
\def\beqn{\begin{eqnarray}}
\def\eeqn{\end{eqnarray}}
\def\sbeqn{\begin{subeqnarray}}
\def\seeqn{\end{subeqnarray}}
\def\BAR{\begin{array}}
\def\EAR{\end{array}}
\newcommand{\lp}{\left(}
\newcommand{\lb}{\left\lbrack}
\newcommand{\la}{\left\{ }
\newcommand{\rp}{\right)}
\newcommand{\rb}{\right\rbrack}
\newcommand{\ra}{\right\} }
\renewcommand{\theequation}{\thesection.\arabic{equation}} %
\newcommand{\bibiteml}[2]{
                    \bibitem{#1}{#2}
                          }
\newcommand{\calle}[1]{(\ref{#1})}
\newcommand{\ket}[1]{\left|\, {#1} \,\right\rangle }
\newcommand{\bfm}{\raisebox{0pt}{{\boldmath$m$\unboldmath}}}
\newcommand{\bfn}{\raisebox{0pt}{{\boldmath$n$\unboldmath}}}
\begin{document}

\rightline{TAUP 2358-96}
\vspace{2cm}

\begin{center}
{\Large\bf   
MINIMAL MODELS OF CFT ON ${\Bbb Z}_N$-SURFACES}
\end{center}

\begin{center}
S. A. Apikyan$^\dagger$\\
Theoretical Physics Department\\
Yerevan Physics Institute\\
 Alikhanyan Br.st.2, Yerevan, 375036 ARMENIA
\end{center}

\vspace{5mm}
\begin{center}
C. J. Efthimiou$^\ddagger$\\
Department of Physics and Astronomy\\
Tel-Aviv University\\
Tel Aviv, 69978 ISRAEL
\end{center}

\vspace{2cm}

\begin{abstract}
The conformal field theory on a ${\Bbb Z}_N$-surface is
studied by mapping it on the branched sphere. Using a
coulomb gas formalism we construct the minimal models 
of the theory.
\end{abstract}

\vfill
\hrule
\ \\
\noindent
$\dagger$ e-mail address: apikyan@vx2.yerphi.am    \\
$\ddagger$ e-mail address: costas@ccsg.tau.ac.il 
\newpage

\section{Introduction}

In recent years there have been many physical applications of the theory
of algebraic surfaces in a wide spectrum of subjects, such as
string theory \cite{strings},
conformal field theory \cite{cfts},
solutions of the Einstein equations \cite{EinsEq},
integrable models \cite{IM},
the theory of defects \cite{defects}.
The best known and simplest
algebraic curves are the hyper-elliptic surfaces
(HESs) \cite{FK};
therefore it is not a
surprise that the majority of the 
references cited above deals with them.
Recently, HESs have also appearred in the Seiberg-Witten theory
of 4-dimensional gauge theories \cite{SW}.

However,
interesting physical applications are not limited to HES
but include other algebraic surfaces as well \cite{other}.
In fact, one would like
to understand quantum field theories on curved space-time
with non-trivial topology.

HESs have a
${\Bbb Z}_2$ automorphism. 
 An immediate generalization is therefore the study of
${\Bbb Z}_N$-surfaces (ZNSs),  i.e. surfaces that will have a 
${\Bbb Z}_N$ automorphism. Although many mathematical aspects
of ZNSs are not so tractable as in the case
of HES, they are still characterized by a high degree of 
symmetry\footnote{Of course, one can find Riemann surfaces with
an even larger symmetry, e.g. ${\Bbb Z}_N\times{\Bbb Z}_M$.
Increasing the symmetry we restrict ourselves to
smaller portions of the moduli space of Riemann surfaces;
nevertheless, surfaces of higher symmetry may be important as well.}
and there are certain physical constructions on them,
such as the Conformal Field Theory (CFT), that can be
studied relatively easy.

The subject of the present paper is to discuss 
minimal models (MMs) of CFT on ZNSs.
Surprizingly, these have not been discussed in the 
existing literature. We thus construct the basic features
of the CFT on the ZNSs, such as the corresponding  graded Virasoro algebra,
the Coulomb Gas Formalism (CGF), the minimal models and their
Kac spectrum. Our construction generalizes the construction
of Crnkovic et all. \cite{cfts} on HESs.
 In a fortcoming publication
\cite{AE2}, we will examine integrable perturbations of
these models. 

Incidentally, we notice that  Krichever and Novikov \cite{KN} have
studied an important  generalization
of the Virazoro algebra (KN algrbra) on general
Riemann surfaces. Their construction has been furthemore
studied and extended by various other
people in several directions. However, the graded algebra \calle{algebra}
of ZNSs
enjoys a particular simple structure, a property that it is not
shared by the general KN algebra.

\section{A primer to 
         ${\Bbb Z}_N$-Symmetric Riemann Surfaces}
\label{ZN-surfaces}

 A ${\Bbb Z}_N$-symmetric Riemann surface $X^{(N)}_g$
is defined as the set of all points $(z,y)$ such that
$y(z)$ satisfies the following equation
\beq
\label{basic-eq}
  y^N(z)=\prod_{i=1}^h\,(z-w_i)^{n_i}~,~~~~~n_i>0~.
\eeq
Notice that, without loss of generality, we can assume that
\beq
\label{range}
    0<n_i<N~,~~~~~\forall i~,
\eeq
 since if $n_{i_0}=q\, N+r$ we can define a new variable
$\zeta= y/(z-w_{i_0})^q$, in terms of which the power $r$ of the 
factor $z-w_{i_0}$ satisfies \calle{range}.
A surface for which all $n_i$ are 1 is called non-singular;
if at least one of them differs from unity, then the surface is called
singular. Obviously, a singular surface can be consindered
a special case of a non-singular one, in the case some
of the branch points $w_i$ coincide.

We  assume that the point at
infinity is not a branch point. This assumption is not
essential but it is introduced to avoid extra (purely technical)
complications. In terms of equations, it implies
\beq
    \sum_i\, n_i = 0\,{\rm mod}N~.
\label{condition1}
\eeq
Another technical assumption is the requirement that the G.C.D.
of $N$ with each of $n_i$ must be 1. In particular, this is true
for the non-singular surfaces or for surfaces with prime $N$.

The genus $g$ of a ${\Bbb Z}_N$-surface can be calculated very easily
using the Riemann-Hurwitz theorem. According to this theorem,
if the function $f:M\rightarrow N$ maps the Riemann surface $M$ of genus $g$
to the Riemann surface $N$ of genus $\gamma$ and the degree of the mapping
is $n$, then
\beq
\label{RH}
   g=n(\gamma-1)+1+{B\over 2}~,
\eeq
where $B$ is the ramification index.
In our case, the function  $z:X_g^{(N)}\rightarrow S^2$, maps the
the  ${\Bbb Z}_N$-surface $X_g^{(N)}$ with genus $g$
onto 2-sphere with $\gamma=0$ such that $n=N$. Now, since every point
of the $S^2$ is covered exactly $N$ times and therefore the branch number is
$N-1$ and since  there are $h$ branch points, we conclude that
$B=h(N-1)$. Sabstituting in equation \calle{RH}, we find
\beq
   g={ (N-1)(h-2)\over 2}~.
\eeq

The ZNS \calle{basic-eq} has $h$ (complex) parameters. Three of them can be
mapped by $SL(2,{\Bbb R})$ invariance 
to $0, 1, \infty$. Therefore, the moduli space ${\cal M}_{\rm ZMS}$
 of ZNSs has dimension
$$
   \dim {\cal M}_{\rm ZMS} =h-3 = {2g\over N-1}-1~.
$$
Comparing to the dimension of the moduli space ${\cal M}$ of generic Riemann
surfaces
$$
  \dim {\cal M} =
  \cases{  1~, & $g=1~,$\cr
        3g-3~, & $g>1~,$\cr }
$$
we conclude that ZNSs do not exchaust all Riemann surfaces. However,
notice that all genus $g=1,2$ are HESs ($N=2$).

The number of independent holomorphic differentials on  $X^{(N)}_g$ equals
the genus $g$. To count them, we write down all such independent
differentials:
\beq
     \Omega_{lj}={z^{j-1}\, dz\over y_l }~,
\eeq
where $l=1,2,\dots,N-1$ and where we introduced the 
quantity\footnote{The symbol $\{ x\}$ denotes the fractional 
part of $x$. If $\lfloor x \rfloor$ is the integer part of
 $x$ then $x=\lfloor x \rfloor + \{ x\}$. Obviously, $0\le
\{ x\} <1$.}
\beq
  y_l= \prod_{i=1}^h\, (z-w_i)^{ \{ {ln_i\over N} \} }~.
\eeq
 The index $j$ counts counts the number of
independent holomorphic differentials for a fixed $l$; in particular
  $j=1,2,\dots,\le j_{\rm max}(l)$ with
\beq
    j_{\rm max}(l)=\cases{ \sum_i\,\la {ln_i\over N}\ra-1~,
          & if $~\sum_i\, \la{ln_i\over N}\ra >1~,$\cr
           0~, & otherwise~. \cr
               }
\eeq
Notice that the genus $g$ of  $X^{(N)}_g$ equals the total number of
independent holomorphic differentials
\beq
   g= \sum_{l=1}^{N-1}\, \lfloor j_{\rm max}(l) \rfloor= 
   { (N-1)(h-2) \over 2} ~,
\eeq
as it should be.

\section{CFT on ${\Bbb Z}_N$-Surfaces}

\subsection{The Algebra}

We label the $N$ sheets of the
Riemann ${\Bbb Z}_N$-surface $X^{(N)}_g$
by the numbers $l=0,1,\dots,N-1$:
\begin{equation}
    y^{(l)}(z)= 
    \omega^l \,\prod_{j=1}^h\, (z-w_j)^{n_j\over N}~,
\end{equation}
where $\omega$ is the $N$-th root of unity
\beq
   \omega = e^{{2\pi i\over N} }~.
\eeq
Let $\{ A_a,\, B_a \}$ be the basic cycles for the
monodromy group of the surface.
 As we encircle the point $w_i$  along the
contours $A_a,\, B_a$,
in the case of an $A_a$ cycle  we stay on the
same sheet, while in the case of a
  $B_a$ cycle we  pass from the $l$-th sheet
 to the $(l+n_i)$-th one.
We shall denote the process of encircling the points $w_i$
on the cycles $A_a, \, B_a$ by the symbols
$\hat{\pi}_{A_a}$, $\hat{\pi}_{B_a}$ respectively. 
Then\footnote{When algebraic operations
are indicated on subscripts or superscripts that appear inside parentheses,
then the result is always meant ${\rm mod}N$.}
\beq
\label{monodr1}
  \hat{\pi}_{A_a}
   y^{(l)} = y^{(l)}~,\quad
  \hat{\pi}_{B_a}
   y^{(l)} = y^{(l+n_i)}~.
\eeq
Here these generators form a group
of monodromy that in our case of $N$-sheeted covering of the sphere coincides
with the ${\Bbb Z}_N$ group:
\beq
  \hat{\pi}_{A_a}^N =
  \hat{\pi}_{B_a}^N = 1~.
\eeq

We consider the energy-momentum tensor $T(z,y)$ which is a single valued
field on the ZNS. Projecting on the plane, it becomes a multi-valued
field
$$
   T^{(l)}(z)  \equiv T(z, y^{(l)}(z))~.
$$
We now view the representation of $T$ 
on each the $N$ sheets, 
  $T^{(l)}(z)$, as a field on the sphere.

The   monodromy properties \calle{monodr1}
along the cycles $A_a,~B_a$ imply that the following
boundary conditions should be satisfied by the energy-momentum
tensor:
\begin{equation}
\label{eqmonT}
\hat{\pi}_{A_{a}}T^{(l)}=T^{(l)} ,\quad \hat{\pi}_{B_{a}}T^{(l)}=T^{(l+1)}~.
\end{equation}
It is convenient to pass to a basis, in which the operators
$\hat{\pi}_{A_a}$, $\hat{\pi}_{B_a}$ are diagonal\footnote{Notice that 
in this equation, as well as \calle{eqmonT}, we assumed that
$n_i=1,~\forall i$. In the general case, one has only to substitute
equation \calle{diagT} with
$$  T_{(k)}=\sum_{l=0}^{N-1}\,c_{kl}\,T^{(l)}~, $$
where $c_{lk}$ are some
constants.}
\beq
\label{diagT}
   T_{(k)}= T^\dagger_{(N-k)}(z)=\sum_{l=0}^{N-1}\,\omega^{-kl}\,T^{(l)}~.
\eeq
Then
\beq
\hat{\pi}_{A_{a}}T_{(k)}=T_{(k)}~,\quad\quad
\hat{\pi}_{B_{a}}T_{(k)}= \omega^k\, T_{(k)}~.
\label{BC1}
\eeq

The corresponding operator product expansions
 (OPEs) of the $T_{(k)}$ fields can be
determined by taking into account the
 OPEs of $T^{(l)}$. On the same 
sheet, the OPEs of the two fields $T^{(l)}(z), T^{(l')}(z),$
 are the same as 
that on the sphere, while on different sheets they do not correlate;
this is a statement of the fact that two points $z,w$ can be close
enough only if they are on the same sheet:
\beq
  T^{(l)}(z)T^{(l')}(w) =\left\lbrack {\hat{c}/2\over (z-w)^4}+
  {2\,T^{(l)}(w)\over (z-w)^2}+
  {\partial_wT^{(l)}(w)\over z-w}\right\rbrack\, \delta^{ll'} + {\rm reg}~. 
\label{OPE1}
\eeq
Thus, in the diagonal
basis the OPEs can be found to be
\beq
T_{(k)}(z)T_{(k')}(w)={c/2\,  \delta_{(k+k'),0}\over (z-w)^4}+
{2\,T_{(k+k')}(w)\over (z-w)^2}+
{\partial_wT_{(k+k')}(w)\over z-w} + {\rm reg}~,
\label{OPE2}
\eeq
where 
$$
  c=N\,\hat{c}~,
$$
 and $\hat{c}$ is the central charge in the OPE 
$T^{(l)}(z)T^{(l)}(w)$.
 It is seen from 
\calle{OPE2} that $T_{(k')},~k'\not=0$ is a
primary field with respect to $T_{(0)}$. 

To write the algebra 
\calle{OPE2} in
the graded form we first notice that the CFT on the sphere should have
$N$ sectors, labeled by an integer $s=0,1,\dots,N-1$.
Consequently, the mode expansion of $T_{(k)}$ is given by
\beq
   T_{(k)}(z)
  =  \sum_{n\in {\Bbb Z}}\, z^{n-2+\{ {ks\over N}\} }\,
   L^{(k)}_{-n-\{ {ks\over N}\} }
    ~.
\eeq
Inverting the above relation we have
\beq
  L^{(k)}_{n-\la{ks\over N}\ra}=\int\,
  {dz\over 2\pi i}\, z^{n+1-\{ {ks\over N}\} }
 \, T_{(k)}(z)~.
\eeq
 Standard calculations
lead to the following algebra for the operators
$L_{n-\{ {ks\over N} \} }^{(k)}$:
\beqn
  \lbrack L^{(k)}_{n-\{ {ks\over N}\} },L^{(k')}_{n'-\{ {k's'\over N}\} }\rbrack
 \!\!\!\! &=& \!\! \!\! \Big(n-n'-\{ {ks\over N}\}+
  \{ {k's'\over N}\} \Big)\,L^{(k+k')}_{n+n'-\{ {ks\over N}\}-
  \{ {k's'\over N}\} }
   \nonumber \\
  &+&  \frac{c}{12}
  \Bigg\lbrack \Big( n-\{\frac{ks}{N}\} \Big)^3-
  \Big(n-\{\frac{ks}{N}\}\Big)\Bigg\rbrack \, 
  \delta_{n+n'-\{ {ks\over N}\} - \{ {k's'\over N}\} ,0}
  \delta^{(k+k'),0}~.~~~~~~~
\label{algebra} 
\eeqn

The operators $\overline{L}^{(k)}_{n-\{ {ks\over N}\} }$
satisfy the same relations and 
$\overline{L}^{(k)}_{n-\{ {ks\over N}\} }$
commute with 
$L^{(k)}_{n-\{ {ks\over N} \} }$.

\subsection{The Representations}

 To describe
the representations of the algebra \calle{algebra},
 it is necessary to consider 
separately the non-twisted  sector  with $k=0$ and 
the twisted sectors 
with $k\not=0$. In order  
to write the $\lbrack V_{\lbrack k\rbrack }\rbrack$ representation of the
algebra \calle{algebra} in a more explicit form, it is convenient to
consider the highest weight states.

(i) In the $k=0$ sector, the
highest weight  state 
$ |\, \vec \Delta_{\lbrack 0\rbrack}\, \rangle\equiv\,
 | \, \Delta^{(0)}_{\lbrack 0\rbrack},
 \Delta^{(1)}_{\lbrack 0\rbrack}, \dots
 \Delta^{(N-1)}_{\lbrack 0\rbrack} \rangle$
is determined with the help of a primary field  $V_{\lbrack 0\rbrack}$
by means of the formula
\begin{equation}
\label{state1}
   | \,\vec\Delta_{\lbrack 0\rbrack}\rangle=V_{\lbrack 0\rbrack}\,
    |\, \emptyset
    \rangle ~.
\end{equation}
Using the definition of vacuum, it is easy to see that
\sbeqn
\slabel{eq1}
 L_0^{(k)}\,|\,\vec\Delta_{\lbrack 0\rbrack}\rangle &=&
 \Delta^{(k)}_{\lbrack 0 \rbrack}
 \, |\, \vec\Delta_{\lbrack 0\rbrack} \rangle~ ,\\
 L_n^{(k)}\, |\, \vec\Delta_{\lbrack 0\rbrack} \rangle&=&0
 \quad n \geq 1~,~~~k=0,1,\dots,N~.
\slabel{eq2}
\seeqn

Thus, the Verma module over the algebra \calle{algebra} in the $k=0$
sector
 is obtained by the action
of any number of $L_{-n}^{(p)}$, $p=0,1,2,\dots,N-1,$
operators with  $n>0$
on the state \calle{state1}.

The first null state is at level 1:
\beq
   \ket{\chi} = \sum_{k=0}^{N-1}\, c_k\, L^{(k)}_{-1}
   \ket{\vec\Delta_{[0]}}~,~~~c_0=1~.
\eeq
Acting with the operators $L^{(N-l)}_1,~l=1,2,\dots,N-1$
on $\ket{\chi}$ we find
\beq
   \sum_{k=0}^{N-1}\, c_k\, \Delta^{(N-l+k)}_{[0]} =0 ~.
\eeq
This system of equations has a non-trivial solution
for the constants $c_k$, only iff
\beq
   \det [ \Delta^{(N-l+k)}_{[0]} ] =0 ~.
\eeq
The last condition has $N$ solutions, with the $l$-th solution
($l=0,1,\dots, N-1$) being
\beq
    \Delta^{(k)}_{[0]} =
    \omega^{-kl} \, \Delta^{(0)}_{[0]}  ~.
\eeq

(ii) In the $k\not=0$  sectors, 
we define the vector of highest weight $|\Delta_{\lbrack k\rbrack}\rangle$
 of the algebra to be\footnote{  This assumes that
$N$ is a prime number since otherwise the product $kp$ may be divisible
by $N$; in this case, we have to take into account
additional zero modes. For example, if $N=4$ and $k=p=2$ then
$\{ {kp\over N}\}=0$  and the primary state in the sector $k=2$
should carry an additional weight. In the following, for simplicity,
we shall always assume that $N$ is a prime number. The generalization
to composite numbers is straightforward.}
\begin{equation}
\label{state2}
  |\, \Delta_{\lbrack k\rbrack} \rangle=
  V_{\lbrack k\rbrack}\,\vline \,\emptyset\rangle~,
\end{equation}
where $V_{\lbrack k \rbrack}$ is a primary field 
with respect to $T$. In analogy with the non-twisted sector we
obtain
\sbeqn
\slabel{eq3}
   L_0^{(0)}\,| \,\Delta_{\lbrack k\rbrack} \rangle &=&
   \Delta_{\lbrack k\rbrack} \,| \,\Delta_{\lbrack k\rbrack} \rangle~,\\
\slabel{eq4}
    L_{n-\{ {kp\over N} \} }^{(p)} \,|\, \Delta_{\lbrack k\rbrack} \rangle&=&0,
   \quad n\ge 1 ~,~~p=0,1,\dots,N-1~.
\seeqn

Thus, the Verma module over the algebra \calle{algebra}
is obtained by the action
of any number of $L_{-m-\{ {kp\over N}\} }^{(k)},~k,p=0,1,\dots,N-1,$ 
operators with  $m> - \{ {kp\over N}\}$
on the state  \calle{state2}. 

We notice that for a twisted primary state $\ket{\Delta_{\lbrack k\rbrack}}$, 
the descendent state $\ket{\Delta'_{\lbrack k\rbrack}}= 
L_{-n-\{ {kk'\over N}\} }^{(k')}
\ket{\Delta_{\lbrack k\rbrack}}$, satisfies
\beq
   L_0^{(0)}\,\ket{\Delta'_{\lbrack k\rbrack}}=
   (\Delta+n+{kk'\over N})\, \ket{\Delta'_{\lbrack k\rbrack}}~.
\eeq
The tower of states in the twisted sectors is seen in the following
table (we have ommitted the primary state for simplicity):
\begin{center}
\begin{tabular}{|l|l|l|l|c|}
\hline
level & $k=1$ sector & $k=2$ sector & $k=3$ sector& \dots \\
\hline
 $\Delta+{1\over N}$ & $L_{-1/N}^{(1)}$ &
$L_{-1/N}^{({N+1\over 2})}$ &
 \dots  & \dots \\
\hline
 $\Delta+{2\over N}$ & $L_{-1/N}^{(1)}L_{-1/N}^{(1)}$,
 $L_{-2/N}^{(2)}$ &
 $L_{-1/N}^{({N+1\over 2})}L_{-1/N}^{({N+1\over 2})}, ~ L_{-2/N}^{(1)}$ 
 & \dots & \dots \\
\hline
 $\Delta+{3\over N}$ & $L_{-1/N}^{(1)}L_{-1/N}^{(1)}L_{-1/N}^{(1)}$,
 $L_{-2/N}^{(2)}L_{-1/N}^{(1)}$, $L_{-3/N}^{(3)}$ & \dots
 & \dots & \dots \\
\hline
 $\Delta+{4\over N}$ & $L_{-1/N}^{(1)}L_{-1/N}^{(1)}L_{-1/N}^{(1)}
L_{-1/N}^{(1)}$,  &
\dots & \dots
&\dots  \\
\phantom{$\Delta+{4\over N}$} &
$L_{-2/N}^{(2)}L_{-1/N}^{(1)}L_{-1/N}^{(1)}$ & & & \\
\phantom{$\Delta+{4\over N}$} &
$L_{-3/N}^{(3)}L_{-1/N}^{(1)}$, $L_{-2/N}^{(2)}L_{-2/N}^{(2)}$ &
& & \\
\hline
\dots & \multicolumn{4}{c|}{\dots \dots \dots} \cr
\hline
\end{tabular} 
\end{center}

For each sector $s$ there is a permutation $P_s$ of $(1,2,\dots, N-1)$,
$k \mapsto P_s(k)$,
such that
\beq
\label{permut}
 \la {P_s(k) s\over N}\ra=
 \la {k \over N}\ra ~.
\eeq
In this way, the previous table can be rewritten more compactely:
\begin{center}
\begin{tabular}{|l|l|}
\hline
level & $s$ sector  \\
\hline
 $\Delta+{1\over N}$ &
     $L_{-1/N}^{(P_s(1))}$ 
\\ \hline
 $\Delta+{2\over N}$ &
     $L_{-1/N}^{(P_s(1))}L_{-1/N}^{(P_s(1))}$,
     $L_{-2/N}^{(P_s(2))}$ 
\\ \hline
 $\Delta+{3\over N}$ &
     $L_{-1/N}^{(P_s(1))}L_{-1/N}^{(P_s(1))}L_{-1/N}^{(P_s(1))}$,
     $L_{-2/N}^{(P_s(2))}L_{-1/N}^{(P_s(1))}$, $L_{-3/N}^{(P_3(3))}$ 
\\ \hline
 $\Delta+{4\over N}$ &
    $L_{-1/N}^{(P_s(1))}L_{-1/N}^{(P_s(1))}L_{-1/N}^{(P_s(1))}
    L_{-1/N}^{(P_s(1))}$,  
   $L_{-2/N}^{(P_s(2))}L_{-1/N}^{(P_s(1))}L_{-1/N}^{(P_s(1))}$, 
   $L_{-3/N}^{(P_s(3))}L_{-1/N}^{(P_s(1))}$, $L_{-2/N}^{(P_s(2))}
   L_{-2/N}^{(P_s(2))}$ 
\\ \hline
  \dots & \dots\dots\dots
\\ \hline
\end{tabular} 
\end{center}

The simplest null state in the $s$-th sector
is
\beq
   \ket{\chi}=L_{-1/N}^{(P_s(1))}\ket{\Delta_{\lbrack s\rbrack}}~.
\eeq
From the fact that $\langle\,\chi\,\ket{\chi}=0$, we derive that
\beq
\label{null1}
  \Delta_{\lbrack s\rbrack}= {c\over 24}\, (1-{1\over N^2})~.
\eeq
Notice that the weight of the lowest null state
 is independent of the sector.
In fact, this result is generic to all null states and it is
seen to be an immediate consequence of \calle{permut} and the
algebra \calle{algebra}.

This can be checked explicitly, by finding the weights 
the next simplest null state:
\beq
  \ket{\psi}=\lp L_{-1/N}^{(P_s(1))}L_{-1/N}^{(P_s(1))}
             +b\, L_{-2/N}^{(P_s(2))} \rp \ket{\Delta_{\lbrack s\rbrack}}~.
\eeq
After some calculations, one finds that 
\sbeqn
  \Delta_{\lbrack s\rbrack}&=& {c\over 24}\, (1-{1\over N^2})~,
\slabel{null2}  \\
  \Delta_{\lbrack s\rbrack}&=& 
   {1\over 16 N}\,
   \lb 5 + {c\over 3}\, (2N^2-5) \pm
   \sqrt{ (1-{c\over N}) (25-{c\over N}) } \rb ~.
\slabel{null3} 
\seeqn
Of course, one of the weights corresponds to the null state at the 
previous level.

\subsection{Coulomb Gas Formalism}

Using reference \cite{FF}, Dotsenko and Fateev \cite{DF}
gave the complete solution for the
minimal model correlation functions on the sphere. They were able to write
down the integral representation for the conformal blocks of 
the chiral vertices
in terms of the correlation functions of 
the vertex operators of a free bosonic
scalar field $\Phi(z,\bar z)=\phi(z)\bar\phi(\bar z)$
coupled to a background charge $\alpha_0$. 
This construction has become known as the Coulomb Gas Formalism
(CGF).
In the present case, this approach is also applicable by
considering  a Coulomb gas for each sheet separately 
but coupled to the same background charge:
\beq
T^{(l)}(z)=-\frac{1}{2}(\partial_z\phi^{(l)})^2(z) + i\alpha_0\partial_z^2
\phi^{(l)}(z)~,
\eeq
where 
\sbeqn
  \langle\phi^{(l)}(z)\phi^{(l')}(w)\rangle=-\delta^{ll'}
  \,\ln(z-w)~,\\
  \hat{\pi}_{A_a}\partial_z\phi^{(l)}=\partial_z\phi^{(l)}~ ,\quad
  \hat{\pi}_{B_a}\partial_z\phi^{(l)}=\partial_z\phi^{(l+1)}~.
\seeqn
In this case $\widehat c=1-12\alpha_0^2$ and thus 
\beq
      c=N\, (1-12\alpha_0^2)~.
\eeq

Passing to the basis which diagonalizes the operators $\hat{\pi}_{A_a}$ ,
$\hat{\pi}_{B_a}$, i.e. 
\sbeqn
    \phi_{(k)}&=&\sum_{l=0}^{N-1}\, \omega^{-kl}\,\phi^{(l)}~,\\
    \langle\phi_{(k)}(z)\phi_{(k')}(w)\rangle&=&-N\,\delta_{(k+k'),0}
     \,\ln(z-w)~,\\
    \hat{\pi}_{A_a}\partial_z\phi_{(k)} = \partial_z\phi_{(k)}~ , &&
    \hat{\pi}_{B_a}\partial_z\phi_{(k)} = \omega^k\,\partial_z\phi_{(k)}~,
\seeqn
we finally obtain the bosonization rule for the operators $T_{(k)}$ in the
diagonal basis
\beq
T_{(k)}=-\frac{1}{2N}\,\sum_{s=0}^{N-1}\,
\partial_z\phi_{(s)}\partial_z\phi_{(k-s)} 
+ i\alpha_0\partial_z^2\phi_{(k)}~.
\eeq

In analogy with the Virasoro algebra, we can expand the fields
$\phi_{(s)}$ in modes
\beq
\partial_z\phi_{(k)}(z)=
\sum_{n\in {\Bbb Z}}\, z^{n-1+\{ {kp\over N}\} }\,
a^{(k)}_{-n-\{ {kp\over N}\} }~.
\eeq
Again, quite easily, we obtain
\beq
\lbrack a^{(k)}_{n-\{ {kp\over N}\} },a^{(k')}_{n'-\{ {k'p'\over N}\} }
\rbrack = N(n-\{ {kp\over N}\} ) \, \delta_{n+n'-\{ {kp\over N}\}
 -\{ {k'p'\over N}\} ,0}\,
\delta^{(k+k'),0}~.
\eeq
Equiped with this result, we can calculate the correlation function
of two bosons in the $l$-th sector:
\beq
\label{ffcorel}
\langle \partial_z\phi_{(k)}(z)\partial_w\phi_{(k')}(w)\rangle_{(l)}
= \delta_{(k+k'),0}\, N\,
{ (1-\{ {kl\over N}\} )\, \lp{w\over z}\rp^{-\{ {kl\over N}\} }
+ \{ {kl\over N}\} \, \lp{w\over z}\rp^{1-\{ {kl\over N}\} }
\over (z-w)^2}~.
\eeq
Using this result, we obtain the expectation value of
the energy-momentum tensor on the $l$-th sector to 
be\footnote{If $N$ is not a prime, let $q(N,l)$ be the G.C.D. of 
    $N$ and $l$ and $N=q\,N'$, $l=q\,l'$. Then we observe that
    $$ 
    {1\over 4}\,
    \sum_{k=0}^{N-1}\lb \{ {kl\over N}\}-\{ {kl\over N}\}^2\rb=
    {q\over 4}\,
    \sum_{k=0}^{N'-1}\lb \{ {k\over N'}\}-\{ {k\over N}\}^2\rb=
    {N^2-q^2\over 24N}~.
    $$
   }
\beq
\label{Texpect}
  \langle T_{(0)}(z) \rangle_{(l)}= {\sum_{k=0}^{N-1}\lb
 \{ {kl\over N}\}-\{ {kl\over N}\}^2 \rb\over 4}\,{1\over z^2}=
{N^2-1\over 24N}\,{1\over z^2}~.
\eeq

To simulate the properties of the different sectors, one defines
the twist fields 
$\Sigma_l(z,\overline z|k)=\sigma_l(z|k)\sigma_l(\overline z|k)$,
$l,k=1,2,\dots,N-1$.
These fields are primary fields
\beq
T_{(0)}(z)\sigma_l(w|k)=
{\Delta_{lk}\over (z-w)^2}\, \sigma_l(w|k)+
{\partial_w\sigma_l(w|k)\over z-w} + {\rm reg}~,
\label{Ts-OPE}
\eeq
and create branch cuts. More precisely,
the twist field
$\sigma_l(z;k)$ react (i.e. has non-trivial OPE)
with $\partial_z\phi_{(k)}(z)$ and
$\partial_z\phi_{(N-k)}(z)$ only:
\beq
\partial_z\phi_{(k)}(z)\sigma_l(w|k)=
(z-w)^{\{ {kl\over N}\} }\,
 \widehat\sigma_l(w|k) + {\rm reg}~.
\label{fs-OPE}
\eeq
From the formulas written above, one can easily see that
\beq
 \Delta_{lk} = 
 {1\over 4}\, \lb \{ {kl\over N}\}-\{ {kl\over N}\}^2 \rb~.
\eeq

In the $k=0$ (non-twisted) sector,  
we now  represent the  primary fields
with charges $\alpha^{(s)}_{\lbrack 0\rbrack},~s=0,1,\dots,
N-1,$ by
\begin{equation}
\label{vertex1}
V_{\lbrack 0\rbrack}(z) = 
e^{i\sum_{s=0}^{N-1}\alpha^{(s)}_{\lbrack 0\rbrack}\phi_{(s)}(z)}~.
\end{equation}
A simple calculation gives
\beq
T_{(k)}(z) V_{\lbrack 0\rbrack}(w)=
{\Delta_{\lbrack 0\rbrack}^{(k)}\,\over (z-w)^2}\, V_{\lbrack 0\rbrack}(w)+
{i\sum_{s=0}^{N-1}\alpha^{(s-k)}_{\lbrack 0\rbrack}\partial_w\phi_{(s)}(w)
\over z-w}\, V_{\lbrack 0\rbrack}(w) + {\rm reg}~,
\label{TV-OPE}
\eeq
where
\beq
\label{weights}
    \Delta^{(k)}_{\lbrack 0\rbrack}\lp \alpha^{(s)}_{\lbrack 0\rbrack}\rp
    \equiv\Delta^{(k)}_{\lbrack 0\rbrack}={N\over 2}\, \sum_{s=0}^{N-1}
  \alpha^{(s-k)}_{\lbrack 0\rbrack}
  \alpha^{(-s)}_{\lbrack 0\rbrack} -N\, \alpha_0\, 
  \alpha^{(-k)}_{\lbrack 0\rbrack}~.
\eeq
So 
\beq
   | \vec \Delta_{\lbrack 0\rbrack} \rangle\equiv  \lim_{z\rightarrow 0}\,
   V_{\lbrack 0\rbrack}(z)
   | \emptyset \rangle~.
\eeq
It is easy to see that for the states $| \vec \Delta_{\lbrack 0\rbrack}
 \rangle$ thus constructed, equations \calle{eq1} and \calle{eq2}
are satisfied. 
We notice the following symmetry
\beq
\label{symmetry1}
 \Delta^{(k)}_{\lbrack 0\rbrack}\left(2\alpha_0\delta^{s0}-
\alpha^{(s)}_{\lbrack 0\rbrack}\right)=
 \Delta^{(k)}_{\lbrack 0\rbrack}\left(\alpha^{(s)}_{\lbrack 0\rbrack}
\right)~.
\eeq

In the $k\not=0$ sectors, we propose
\beq
V_{\lbrack k\rbrack}(z)= e^{i\alpha_{\lbrack k\rbrack}\phi_{(0)}(z)}\,
 \sigma_k(z|1)
 \sigma_k(z|2)\dots
 \sigma_k(z|N-1)
  ~.
\eeq
The product of the twist fields is implied by the form of the 
energy-momentum tensor
\beq
 T_{(0)} = -{1\over 2N}\, \partial_z\phi_{(0)}\partial_z\phi_{(0)}
           +i\alpha_0\partial_z^2\phi_{(0)}
           -{1\over 2N}\,
          \sum_{s\not=0}\, \partial_z\phi_{(s)}\partial_z\phi_{(N-s)}
   ~.
\eeq
In representation theory, the first two terms (untwisted boson)
requires the vertex operator, while each additional term introduces
its own twist field.

Defining 
\beq
   | \vec \Delta_{\lbrack k\rbrack} \rangle\equiv  \lim_{z\rightarrow 0}\,
   V_{\lbrack k\rbrack}(z)
   | \emptyset \rangle~,
\eeq
equations \calle{eq3} and \calle{eq4}
are satisfied with
\beq
  \Delta_{\lbrack k\rbrack}\lp \alpha_{\lbrack k\rbrack}\rp
 \equiv\Delta_{\lbrack k\rbrack}=
  {N\over 2}\,  \alpha_{\lbrack k\rbrack}^2-
N\,\alpha_0\,\alpha_{\lbrack k\rbrack}+
{N^2-1\over 24\, N}~.
\eeq
This weight has also the symmetry
\beq
\label{symmetry2}
 \Delta_{\lbrack k\rbrack}\lp 2\alpha_0- \alpha_{\lbrack k\rbrack}\rp=
 \Delta_{\lbrack k\rbrack}\lp \alpha_{\lbrack k\rbrack}\rp~.
\eeq

\subsection{The Screening Charges}

We can easily construct screening charges.
In the $k=0$ sector, a  screening
charge must have singular terms with all 
$T_{(k)}$ which can be exressed as a total
derivative, i.e. 
\beq
\label{charge}
    Q=\int\,{dw\over 2\pi i}\,S(w)
\eeq
 is a screening charge iff
\beq
   T_{(k)}(z)S(w)= {\rm const}~ \partial_w\lp S(w)\over z-w\rp
   +{\rm reg}~, \quad\quad \forall k~,
\eeq
since in this case
\beq
    \lbrack  T_{(k)}(z), Q\rbrack = \oint\,{dw\over 2\pi i}
    \,T_{(k)}(z)S(w)=0~.
\eeq
This will be true if
\beq
  \sum_{s=0}^{N-1}\,\alpha^{(s-k)}_{\lbrack 0\rbrack}\partial_w\phi_{(s)}(w)
   = \Delta^{(k)}_{\lbrack 0\rbrack}\,
  \sum_{s=0}^{N-1}\alpha^{(s)}_{\lbrack 0\rbrack}\partial_w\phi_{(s)}(w)~,
\eeq
or 
\beq
\label{equation}
  \sum_{s=0}^{N-1}\,\la
   \alpha^{(s-k)}_{\lbrack 0\rbrack} - \Delta^{(k)}_{\lbrack 0\rbrack}\,
  \alpha^{(s)}_{\lbrack 0\rbrack} \ra\,
 \partial_w\phi_{(s)}(w)=0~.
\eeq
The solutions to equation \calle{equation} can be found easily:
\beq 
     \Delta^{(k)}_{\lbrack 0\rbrack}= \omega^{-kl}~,
     \quad\quad
     \alpha^{(s)}_{\lbrack 0\rbrack}= \alpha\,\omega^{sl}~,
\eeq
where $l=0,1,\dots,N-1$ and
 $\alpha$ is a constant to be determined by \calle{weights}.
In particular, equation \calle{weights} reduces to the
quadratic  equation
\beq
{N^2\over 2}\,\alpha^2 - N\,\alpha_0\,\alpha -1=0~,
\eeq
which has the two solutions
\beq
\label{a-eq}
   \alpha_\pm(N)\equiv
\alpha_\pm={\alpha_0\over N}\pm {1\over N}\, \sqrt{\alpha_0^2+2}~.
\eeq
For later convenience, let us note the identities
\beq
  \alpha_+\,\alpha_-=-{2\over N^2}~,~~~~~
  \alpha_+ + \alpha_-={2\alpha_0\over N}~.
\eeq
Therefore we conclude that we have the following $2N$ 
screening vertices
\beq
   S_l^\pm(z)= e^{i\alpha_\pm\sum_{s=0}^{N-1}\omega^{sl}\phi_{(s)}(z)}~,
\eeq
defing the charges $Q^\pm_l$ by \calle{charge}.

Of course,
in the $k\not=0$ sectors we must only screen
the charge  for the zeroth field; the corresponding screening charges
$Q,^+, Q^-$ arise essentially as  products of $Q^+_l, Q^-_l$.

\subsection{Null States}   
   
Having the screening charges we can construct many null states
\cite{FZ1}.
Let
\beqn
  |\chi_l^\pm\rangle &=& Q_l^\pm|\vec \Delta_{\lbrack 0\rbrack}
  (2\alpha_0\delta^{s0}-\alpha^{(s)}-\alpha_\pm\omega^{sl}) \rangle
  \nonumber \\
 &=& \oint_{{\cal C}}\, {dz\over 2\pi i}\, S^\pm_l(z)
   \,e^{i\sum_{s}\lbrack 2\alpha_0\delta^{s0}-\alpha^{(s)}-
   n\alpha_\pm\omega^{sl})
 \rbrack \phi_{(s)}(0)}|\emptyset\rangle
  \nonumber\\
 &=& \oint_{{\cal C}}\, {dz\over 2\pi i}\,
 :e^{i\alpha_\pm\sum_s\omega^{sl}\phi_{(s)}(z)}:\,
 :e^{i\sum_{s}\lbrack 2\alpha_0\delta^{s0}-\alpha^{(s)}-
   n\alpha_\pm\omega^{sl})
 \rbrack \phi_{(s)}(0)}:|\emptyset\rangle
  \nonumber
\eeqn
for some closed  contour ${\cal C}$. 
We can rewrite the last integral in the form
$$
|\chi_l^\pm\rangle =
   \oint\, {dz\over 2\pi i}\, z^{N\alpha_\pm
  \lbrack2\alpha_0-\sum_s\alpha^{(s)}\omega^{-sl}-
  N\alpha_\pm\rbrack}\,
    :e^{i\sum_s\la\alpha_\pm\omega^{sl}\phi_{(s)}(z)+
    \lb 2\alpha_0\delta^{s0}-\alpha^{(s)}-\alpha_\pm\omega^{sl}\rb
  \phi_{(s)}(0)\ra}:|\emptyset\rangle~.
$$
The above integral will be well defined and
non-vanishing  iff
\beq
\label{condition}
 N\alpha_\pm \,
  \lbrack2\alpha_0-\sum_s\alpha^{(s)}\omega^{-sl}-
  N\alpha_\pm\rbrack
 =-n_l-1~,
\eeq
where $n_l=0,1,2,\dots$ .
In particular, this condition implies
\beq
 |\chi_l^\pm\rangle\propto \partial_z^{n_l}
    :e^{i\sum_s\la 2\alpha_0\delta^{s0}-
    \alpha^{(s)}\ra \phi_{(s)}(0)}:|\emptyset\rangle~.
\eeq
This  clearly shows that
 $|\chi_l^\pm\rangle$ is a descendant of the state
 $|\vec\Delta_{\lbrack 0\rbrack}(2\alpha_0\delta^{s0}-
\alpha^{(s)})\rangle=|\vec\Delta_{\lbrack 0\rbrack}(\alpha^{(s)})\rangle$.
Since it is also a highest weight state, it must be a null vector.
Notice that for $n_l=0$ we obtain the state $|\vec\Delta_{\lbrack 0\rbrack}(
\alpha^{(s)})\rangle$ and therefore we shall ignore this value.

One can consider the more general case 
\beqn
  |\chi_l^\pm\rangle &=& (Q_l^\pm)^n |\vec \Delta_{\lbrack 0\rbrack}
  (2\alpha_0\delta^{s0}-\alpha^{(s)}-n\alpha_\pm\omega^{sl}) \rangle
  \nonumber \\
 &=& \oint_{{\cal C}_1}\, {dz_1\over 2\pi i}\, S^\pm_l(z_1)
     \oint_{{\cal C}_2}\, {dz_2\over 2\pi i}\, S^\pm_l(z_2)
      \dots
     \oint_{{\cal C}_n}\, {dz_n\over 2\pi i}\, S^\pm_l(z_n)
   \,e^{i\sum_{s}\lbrack 2\alpha_0\delta^{s0}-\alpha^{(s)}-
   n\alpha_\pm\omega^{sl})
 \rbrack \phi_{(s)}(0)}|\emptyset\rangle~,
  \nonumber
\eeqn
where ${\cal C}_1,{\cal C}_2,\dots,{\cal C}_n$ are some closed
contours.
This integral will be now well defined and 
non-vanishing  iff
\beq
\label{condition2}
 (n-1)\,\alpha_\pm^2 N^2+N\alpha_\pm \,
  \lbrack2\alpha_0-\sum_s\alpha^{(s)}\omega^{-sl}-
 n N\alpha_\pm\rbrack
 =-n_l-1~,
\eeq
where $n_l=0,1,2,\dots$. The integer $n$ cancels
in the above equation, the solution being
\beq
 \alpha^{(s)}=\alpha_\pm \,\sum_{l=0}^{N-1}\,
  {1-n_l\over 2}\omega^{sl}~.
\eeq
Therefore, the most general result is
\beq
\label{a0}
 \alpha^{(s)}_{\lbrack 0\rbrack}
 \lb\matrix{n_0&m_0\cr n_1&m_1\cr \dots&\dots\cr n_{N-1}&m_{N-1}\cr}\rb
 =\lp \sum_{l=0}^{N-1}\,
  {1-n_l\over 2}\omega^{sl}\rp \alpha_+ +
  \lp \sum_{l=0}^{N-1}\,
  {1-m_l\over 2}\omega^{sl}\rp \alpha_-~,
\eeq
where $n_l, m_l$ take the values $1,2,\dots$ .
The corresponding weights are
\beqn
\label{D0}
 \Delta^{(s)}_{\lbrack 0\rbrack}
 \lb\matrix{n_0&m_0\cr n_1&m_1\cr \dots&\dots\cr n_{N-1}&m_{N-1}\cr}\rb
 &=& {N^2\over 2}\,\sum_{l=0}^{N-1}\,\lp
  {1-n_l\over 2}\, \alpha_+ +
  {1-m_l\over 2}\, \alpha_- \rp^2\,\omega^{-sl} 
 \nonumber \\ &-&
  N\alpha_0\, \sum_{l=0}^{N-1}\,\lp {1-n_l\over 2}\, \alpha_+ +
  {1-m_l\over 2}\, \alpha_- \rp \,\omega^{-sl}~.
\eeqn
Formulae \calle{a0} and \calle{D0} consist Kac's spectrum in the
$k=0$ sector.
Notice that the weight $\Delta^{(0)}_{\lb 0\rb}$ corresponding
to the actual energy-momentum tensor is real, while the
rest weights are in general complex. However, they satisfy
the expected condition
\beq
    \Delta^{(s)*}_{\lbrack 0\rbrack}=  
    \Delta^{(N-s)}_{\lbrack 0\rbrack}
\eeq

Similarly, one can check that  the most general result for
the null states in the $k\not=0$ sectors is given by
\beq
\label{ak}
 \alpha_{\lbrack k\rbrack}
 \lb n\, m\rb
  ={N-n\over 2} \alpha_+ +
  {N-m\over 2} \alpha_-~,
\eeq
where $n, m$ take the values $1,2,\dots~$.
The corresponding weights are
\beq
\label{Dk}
 \Delta_{\lbrack k\rbrack}
 \lb n\, m \rb
 = {N\over 8}\,\lb (n\alpha_+ +m \alpha_-)^2 - N^2(\alpha_+ +\alpha_-)^2
  \rb +{N^2-1\over 24\, N}~. 
\eeq
Formulae \calle{ak} and \calle{Dk} consist Kac's spectrum in the
$k\not=0$ sector.

\subsection{Correlation Functions and Fusion Rules} 

The $L$-leg correlation function
\beq
 \langle\,
 \Phi_1\lb\bfn_1\, \bfm_1\rb
 \Phi_2\lb\bfn_2\, \bfm_2\rb
 \dots
 \Phi_L\lb\bfn_L\, \bfm_L\rb
 \,\rangle 
\eeq
of the $L$-fields in the $k=0$ sector 
will be non-vanishing if there are natural numbers $M_l, N_l$ such that the
correlation function
\beq
 \langle\, V_{\lb 0\rb}(\vec\alpha_1)
           V_{\lb 0\rb}(\vec\alpha_2)
           \dots
           V_{\lb 0\rb}(\vec\alpha_L) 
      \prod_l\, (Q_l^+)^{M_l}
      \prod_l\, (Q_l^-)^{N_l}
    \rangle
\eeq
is non-vanishing.

The non-vanishing 3-leg correlation functions
will provide the fusion rules. In particular, 
the field $\Phi\lb\bfn\, \bfm\rb$
will appear in the OPE of $\Phi\lb\bfn'\, \bfm'\rb$
and $\Phi\lb\bfn''\,\bfm''\rb$  under the conditions
the functions
\sbeqn
\slabel{3leg-1}
    \langle\, V_{\lb 0\rb}(\vec\alpha^*)
           V_{\lb 0\rb}(\vec\alpha')
           V_{\lb 0\rb}(\vec\alpha'')
      \prod_l\, (Q_l^+)^{M_l}
      \prod_l\, (Q_l^-)^{N_l}
    \rangle \\
\slabel{3leg-2}
    \langle\, V_{\lb 0\rb}(\vec\alpha)
           V_{\lb 0\rb}(\vec\alpha^{\prime*})
           V_{\lb 0\rb}(\vec\alpha'')
      \prod_l\, (Q_l^+)^{M_l}
      \prod_l\, (Q_l^-)^{N_l}
    \rangle \\
\slabel{3leg-3}
    \langle\, V_{\lb 0\rb}(\vec\alpha)
           V_{\lb 0\rb}(\vec\alpha)
           V_{\lb 0\rb}(\vec\alpha^{\prime\prime*})
      \prod_l\, (Q_l^+)^{M_l}
      \prod_l\, (Q_l^-)^{N_l}
    \rangle 
\seeqn
are non-vanishing.
For \calle{3leg-1}, charge conservation implies that
\beq
  0=\sum_l\,\omega^{sl}\,\lb \alpha_+\lp
   N_l +{1-n'_l-n''_l+n_l\over2}\rp +
   \lp M_l +{1-m'_l-m''_l+m_l\over2}\rp \rb~.
\eeq
From these equations, $s=0,1,\dots,N-1$, we find
\sbeqn
    N_l +{1-n'_l-n''_l+n_l\over2}&=&0~,\\
    M_l +{1-m'_l-m''_l+m_l\over2}&=&0~.
\seeqn
Equivalently
\sbeqn
\label{ineq1}
    n_l\le n'_l+n''_l-1~,
    \quad\quad
    -n'_l-n''_l+n_l={\rm odd}~,\\
    m_l\le m'_l+m''_l-1~,
   \quad\quad
   -m'_l-m''_l+m_l={\rm odd}~.
\seeqn
In the same way, equation \calle{3leg-2} gives
\sbeqn
\label{ineq2}
    n'_l\le n_l+n''_l-1~,
    \quad\quad
    +n'_l-n''_l-n_l={\rm odd}~,\\
    m'_l\le m_l+m''_l-1~,
   \quad\quad
   +m'_l-m''_l-m_l={\rm odd}~,
\seeqn
and equation \calle{3leg-3}
\sbeqn
\label{ineq3}
    n''_l\le n'_l+n_l-1~,
    \quad\quad
    -n'_l+n''_l-n_l={\rm odd}~,\\
    m''_l\le m'_l+m_l-1~,
   \quad\quad
   -m'_l+m''_l-m_l={\rm odd}~.
\seeqn
Taking into account all previous results,
the OPEs are
\beqn
\Phi\lb\bfn'\, \bfm'\rb
\Phi\lb\bfn''\, \bfm''\rb &=& 
\sum_{\scriptsize
\BAR{c}n_0=|n'_0-n''_0|+1\\ n_0+n'_0+n''_0={\rm odd}\EAR}^{n'_0+n''_0-1}\,
\dots
\sum_{\scriptsize
\BAR{c}n_{N-1}=|n'_{N-1}-n''_{N-1}|+1\\ n_{N-1}+n'_{N-1}+n''_{N-1}=
{\rm odd}\EAR}^{n'_{N-1}+n''_{N-1}-1} \nonumber \\
&& \sum_{\scriptsize
\BAR{c}m_0=|m'_0-m''_0|+1\\ m_0+m'_0+m''_0={\rm odd}\EAR}^{m'_0+m''_0-1}\,
\dots
\sum_{\scriptsize
\BAR{c}m_{N-1}=|m'_{N-1}-m''_{N-1}|+1\\ m_{N-1}+m'_{N-1}+m''_{N-1}=
{\rm odd}\EAR}^{m'_{N-1}+m''_{N-1}-1}\,
\Phi\lb\bfn\, \bfm\rb~.~~~~~~~~~~
\eeqn

Now, let us introduce the symbol
\beq
 \sigma_{(k)}(z)
 =\sigma_k(z|1)
 \sigma_k(z|2)\dots
 \sigma_k(z|N-1)~.
\eeq
These fields satisfy  the following algebra:
\sbeqn
 \sigma_{(k)}(z)
 \sigma_{(l)}(w) &=& 
 {\sigma_{(k+l)}(z)\over (z-w)^\Delta}+\dots~,
 ~~~~~ k+l\not=N~,
 \\
 \sigma_{(k)}(z)
 \sigma_{(N-k)}(w) &=& 
 \sum \, C_{\vec p}\,
 (z-w)^v\, e^{i\vec p \vec\phi(w)}~.
\seeqn
Using the screening charges, one can now write down the fusion
rules among any fields in the theory.

\subsection{Minimal Models}

In the case 
\beq
  \alpha_+={\sqrt{2}\over N}\, \sqrt{p\over q}~,
  \quad\quad
  \alpha_-=-{\sqrt{2}\over N}\, \sqrt{q\over p}~,
\eeq
where $q,p$ are two relatively prime natural numbers,
we notice the following symmetry
$$
 \Delta^{(k)}_{\lbrack 0\rbrack}
 \lb\matrix{n_0&m_0\cr n_1&m_1\cr \dots&\dots\cr n_{N-1}&m_{N-1}\cr}\rb
 = 
  \Delta^{(k)}_{\lbrack 0\rbrack}
 \lb\matrix{q-n_0&p-m_0\cr n_1&m_1\cr \dots&\dots\cr n_{N-1}&m_{N-1}\cr}\rb
  = \dots
  \Delta^{(k)}_{\lbrack 0\rbrack}
 \lb\matrix{n_0&m_0\cr n_1&m_1\cr \dots&\dots\cr q-n_{N-1}&p-m_{N-1}\cr}\rb
  ~,
$$
in the untwisted sector and
$$
  \Delta^{(k)}_{\lbrack s\rbrack}[n~m]
 =
  \Delta^{(k)}_{\lbrack s\rbrack}[q-n ~p-m]~,
$$
in the twisted sectors.
Therefore the corrresponding fields have to be identified.
In this case the OPEs close with a finite number of fields
\sbeqn
  1\le n_l \le q~,\quad\quad
  1\le m_l \le p~,\quad\quad s=0~,
  \\
  1\le n \le q~,\quad\quad
  1\le m \le p~,\quad\quad s\ne 0~.
\seeqn
We call the corresponding models the minimal models. Among them,
there is a series of models $q=p+1$ which is also unitary.
In this case, the central charge is
\beq
\label{cunitary}
   c=N\lp 1-{6\over p(p+1)}\rp~.
\eeq
Unitary representations can be constructed using the
coset construction \cite{GKO} (also see the recent review see \cite{HKOC})
 based on the group
$$
 G=\underbrace{SU(2)\times SU(2)\times\dots\times SU(2)}_N~,
$$
i.e. forming the coset
\beq
     {\widehat G_{p-2} \times \widehat G_1\over \widehat G_{p-1} }
\eeq
at the indicated levels (for $SU(2)$),
 one recovers the central charge given
by equation \calle{cunitary}.

\vspace{1cm}
\noindent {\large\bf Aknowledgements}

We thank Professor L. Takhtajan for reading the manuscript.
S. A. would like to thank the High Energy Group of 
Tel Aviv University for its hospitality.

\end{document}